\documentclass[11pt]{article}
\usepackage{graphicx,caption2,subfigure}
\makeatletter
\newcommand{\singlespacing}{\let\CS=\@currsize\renewcommand{\baselinestretch}{1.0}\tiny\CS}
\newcommand{\doublespacing}{\let\CS=\@currsize\renewcommand{\baselinestretch}{1.5}\tiny\CS}

\begin{document}
\title {On some properties of Deuteron and Antideuteron production in high energy lead-lead collisions at 158A GeV}
\author { Goutam
Sau$^1$\thanks{e-mail: sau$\_$goutam@yahoo.com} $\&$ S.
Bhattacharyya$^2$\thanks{e-mail: bsubrata@www.isical.ac.in
(Communicating Author)}\\
{\small $^1$ Beramara RamChandrapur High School,}\\
 {\small South 24-Pgs, 743609(WB), India.}\\
 {\small $^2$ Physics and Applied Mathematics Unit(PAMU),}\\
 {\small Indian Statistical Institute, Kolkata - 700108, India.}}
\date{}
\maketitle
\bigskip
\begin{abstract}
We would attempt here to understand some properties of the
transverse momentum ($p_T$) and rapidity ($y$) spectra on production
of deuteron ($d$) and antideuteron($\overline{d}$) in lead-lead
$(Pb+Pb)$ collisions at 158A GeV recently reported by NA49
collaboration. Starting from some basic properties of $p+p$
reactions for production of secondary proton-antiprotons the cases
of production of the composite set of particles, like $d$ and
$\overline{d}$, would be analysed. Some ratio-behaviours would also
be dealt with in the light of the same approaches. It is found that
the combination of the models put into use here capture modestly
well the trends of the data on some important observables. Some
limitations of the approach would also be pointed out in the end.
\end{abstract}

\newpage
\singlespacing
\section{Introduction}
Studies on baryon-antibaryons and light composite cluster
formation\cite{Anticic1}-\cite{Ma1} constitute now a very important
corner from both theoretical and experimental viewpoints comprising
particle physics and astroparticle physics aspects. The observations
and measurements on deuteron and antideuteron production in
relativistic heavy-ion collisions, it is believed, could probe the
later stages of the evolution of hypothesized new state of strongly
interacting matter, called Quark-Gluon Plasma
(sQGP)\cite{Shuryak1}-\cite{Tannenbaum1}. It is also taken for
granted that after the initial expansion of QGP matter and
subsequent cooling, nucleons (antinucleons) in spatial proximity and
with neighbouring momenta might coalesce to form light nucleus
(antinucleus) clusters. The sensitivity of light-nucleus production
to the space-time evolution of the interaction region and collision
dynamics imparts such studies on them a special degree of
importance, and render them quite relevant.
\par
The interests in such studies on light cluster formation particle
like deuteron-antideuteron, triton-antitriton, helium-antihelium etc
sping, in the main, from (i) the prediction of excess
antibaryon\cite{Ellis1,Heinz1} (antimatter) production in
central nucleus-nucleus collisions relative to $p+p$ in the QGP
phase. It is claimed that if antibaryon abundances are not in
equilibrium or not strongly affected by annihilation after chemical
freeze-out some of the initial enhancement may survive till the
final stage of the collision process at the end of which secondary
particles begin to emanate; (ii) the persistent controversies around
the observations of excess of matter over antimatter against the
general background of matter-antimatter
asymmetry\cite{Canetti1}-\cite{Basini1}.
\par
In the present work we primarily concern ourselves with the
understanding of some features of very recent experiments on $d$,
$\overline{d}$ production at various centrality by NA49
collaboration in high energy $Pb+Pb$\cite{Anticic1} collisions. This
study by NA49 collaboration is the first report that came to light
with measurements on $d$ and $\overline{d}$ at various centralities
in high energy $Pb+Pb$ collisions, for which this measurement has
assumed high degree of prominence. And just because of it, we have
been attracted to this fresh bid of study on $d$ and $\overline{d}$
production in $Pb+Pb$ collisions at 158A GeV.
\par The outline of
our approach is as follows. Against the general background of what
is known as the coalescence picture, the explanation is attempted
with the help of a new combination of models (NCM), of which the
first one is for nucleon-nucleon collision and the other one is for
nucleus-nucleus interactions as is detailed in the next paragraph.
Besides, a new parametrization for the nature of mass number (A)
dependence for nucleus-nucleus interactions is introduced. And by
making use of all of them we will first try here to explain the
nature of $p_T$-spectra on $d$ and $\overline{d}$ production in
nucleus-nucleus collisions. Thereafter, we will attempt to
interpret, in addition, characteristic properties of the newly
obtained data on rapidity spectra of $d$ and $\overline{d}$ in the
same reactions at relatively high energies.
\par
The model for nucleon-nucleon interactions used here was formulated
and forwarded long ago by Hagedorn\cite{Hagedorn1} which has been
extensively used by us in interpreting the various
aspects\cite{De1,Bhattacharyya2} of high energy particle and nuclear
collisions. Thus, our next objective here is to check whether
Hagedorn's Model (HM) can also be used for interpreting the
deuteron-antideuteron production phenomena, and which would, in
turn, also help us to check whether the final working formula
utilized here for building up of a straight corridor between
nucleon-nucleon and nucleus-nucleus collisions would be sufficiently
effective or not. However, the antideuteron data are too sparse and
they suffer from high degree of measuremental uncertainties with
large error bars for which they hardly suffice to make the study
very convincing.
\par
Some other preemptive comments are in order here. First, in the
present work, we are not going to offer any new insights into or any
refinements of the used coalescence picture. Second, experimental
data points on the coalescence factor have been used here as an
indirect way to assess the merit and utility of the NCM. Besides,
the combination of the models that are used here also attempts to
explain the features of transverse momenta spectra of antiprotons
and antideuterons in nucleon-nucleon collisions at high energies,
and some related features of antideuteron production as well. Third,
we attempt to interpret the available limited data on
rapidity-dependence properties of the same species ($d$ \&
$\overline{d}$) with the same combinational approach. Fourth and
final, we are certainly not going to dwell upon the present issues
of antideuteron production against the broad background of any
general matter-antimatter controversy, because this is beyond the
purview of the present study.
\par
The plan of the paper is as follows : In Section 2 we give a brief
outline of the basic Coalescence Approach (CA). Section 3 and
Section 4 provides the tools, that is, the synopses of the models
which are put into use here as the two principal concerns of this
study. In Section 5 we summarise the calculational results based on
the models of our choice. The last section is for final discussion
and conclusion.

\section{The Coalescence Model : In a Nutshell}
It is well known that the studies on deuteron production are
commonly grounded on the tacit acceptance of the coalescence
picture. According to this view, deuteron production with a certain
velocity is proportional to the number of antiprotons and neutrons
that have similar velocities; and the coalescence factor is
contingent upon the distribution of nucleons. This property guides
us to determine the source size of the nucleon from the ratio
defined, one assumes, by the form\cite{Dasgupta1,Mekjian1} :
\begin{equation}
B_2(p) = \frac{E_d \frac{d^3N_d}{dP^3}}{(E_p \frac{d^3N_p}{dp^3})^2}
\end{equation}
Wherein $B_2(p)$ is termed the `coalescence parameter' which is
inversely proportional to the volume of the particle source, and
wherein the yield of neutrons is supposed to be equal to that of
protons. The deuteron momentum $P$ is twice the antiproton momentum
$p$. Measurement of neutrons is normally avoided. The factor in
subscript of $B$ represents simply the fact whether
deuteron/antideuteron or any other antiparticle-particle pair is
being studied.
\section{Present Approach : A Brief Outline of the New Combinational Model (NCM)}
This subsection dwells upon the adopted methodology for $p_T$
studies of $p(\overline{p})$ or $d(\overline{d})$. The generalized
form of the inclusive cross-section for production of either
antiproton or antideuteron is taken to be represented here by
\begin{equation}
E\frac{d^3\sigma}{dp^3}\mid_{P+P \rightarrow Q+X} = C_1(1+\frac{p_T}{p_0})^{-n}
\end{equation}
where $Q$ stands for the secondary particle produced in any specific
collisions, $p_T$ is the transverse momentum of $Q$, and $C_1$,
$p_0$, $n$ are the constants. The above form presented by expression
(2) is an obvious adaptation of HM\cite{Hagedorn1} for particle
production in nucleon-nucleon collisions.
\par
But in the present study we are interested to pursue a particular
aspect of the nucleus (A)-nucleus (B) collisions, for which we try
to build up a bridge or linkage between nucleon-nucleon ($p+p$) and
nucleus-nucleus ($A+B$) collisions. With a view to obtaining such a
connective route, let us propose here a form as was prescribed first
by Peitzmann\cite{Peitzmann1} which we refer to as Peitzmann's
Approach (PA) and was utilized by us in some previous work\cite{De2,
De5} :
\begin{equation}
E\frac{d^3\sigma}{dp^3}\mid_{A+B \rightarrow Q+X} \sim (A.B)^{f(p_T)}E\frac{d^3\sigma}{dp^3}\mid_{P+P \rightarrow Q+X}
\end{equation}
with the following subsidiary set of relations
\begin{equation}
f(p_T) = (1 + \alpha p_T - \beta p_T^2),
\end{equation}
\begin{equation}
E\frac{d^3N}{dp^3} = \frac{1}{\sigma_{in}} E\frac{d^3\sigma}{dp^3}
\end{equation}
where $\sigma_{in}$ is the inelastic cross-section, A \& B are mass
numbers for the colliding nuclei and $\alpha$ \& $\beta$ are the
coefficients to be chosen separately for each A+B collisions (and
also for A+A collisions when the projectile and the target are
same).
\par
Using all the expressions from Eq. (2) to Eq. (5), one obtains
finally,
\begin{equation}
E\frac{d^3N}{dp^3}\mid_{A+B \rightarrow Q+X} = C_2(A.B)^{(1+\alpha p_T-\beta p_T^2)}(1+\frac{p_T}{p_0})^{-n}
\end{equation}
where $C_2$ is the normalization term which includes function(s) of
rapidity or rapidity density for the specific $A+B \rightarrow Q+X$
process. By our ascription of the form $f(p_T)$ given in Eq. (4) we
introduce first what is called here De-Bhattacharyya Parametrization
(DBP). The choice of this form is not altogether a sheer
coincidence. In dealing with the EMC effect \textbf{[}\emph{The EMC
effect is just a departure of the ratio of the measured structure
function of the nucleons (of deuterium and iron nuclei) through deep
inelastic $\mu\mu'$ (projectile muon and scattered muon, while the
target is the nucleon - deuterium or iron) and $ee'$ (projectile
electron and scattered electron, while the target is the nucleon -
deuterium or iron) scattering}\cite{Bhattacharyya1}-\cite{Bodek1}.
\emph{The ratios, while expected to rise with an increasing values
of a scaling variable (say, x) very slowly the results depicted
first clear diminishing trends with x; and then with further
increasing values of the same the ratio began to rise modestly
prominently. One of the authors here (SB) attempted to explain quite
successfully this `anomaly' with a polynomial nature of A-dependence
with the same `x'. Thus, the implication of the EMC effect is : the
nucleus viewed as a collectivity of nucleons behave distinctly
differently vis-a-vis the scattering processes from a single
composed nucleus. This message was and is of high physical import in
the realm of both particle and nuclear physics.}\textbf{]} in the
lepton-nucleus collisions; the clue devised by
Bhattacharyya\cite{Bhattacharyya1} that resolved the complex nature
of A-dependence of the ratio stimulated us to make a similar choice
with both the $p_T$ and $y(\eta)$ variables. In recent times, this
parametrization (DBP) is being applied by our group to interpret the
measured data on the various aspects\cite{De2, De5} of
particle-nucleus and nucleus-nucleus interactions at high energies.
In the recent past Hwa et al.\cite{Hwa1} also made use of this sort
of relationship in a somewhat different context. The underlying
physics implication of this parametrization has to stem mainly from
the expression (6) which can be identified as a clear mechanism for
the switch-over of results obtained for nucleon-nucleon $(p+p)$
collision to those for nucleus-nucleus interactions at high energies
in a direct and straightforward manner. The polynomial exponent of
the product term on $A+B$ takes care of the totality of the nuclear
effects.
\par
The physical foundation that has been attempted to be built up is
inspired by thermodynamic pictures, whereas the quantitative
calculations are based on a sort of pQCD-motivated power-law formula
represented by Eq. (2). This seems to be somewhat paradoxical,
because it would be hard to justify the hypothesis of local thermal
equilibrium in multihadron systems produced by high energy
collisions in terms of successive collision of the QCD-partons (like
quarks and gluons) excited or created in the course of the overall
process. Except exclusively for central heavy ion collisions, a
typical parton can only undergo very few interactions before the
final-state hadrons "freeze-out", i.e. escape as free particles or
resonances. The fact is the hadronic system, before the freeze-out
starts, expands a great deal - both longitudinally and transversally
- while these very few interactions take place\cite{VanHove1}. But
the number of parton interactions is just one of the several other
relevant factors for the formation of local equilibrium. Of equal
importance is the parton distribution produced early in the
collision process. This early distribution is supposed to be a
superposition of collective flow and highly randomized internal
motions in each space cell which helps the system to achieve a
situation close to the equilibrium leading to the appropriate values
of collective variables including concerned and/or almost concerned
quantities. Further, one would note that the approach used in the
present paper is an effective parametrization which is not
rigorously derived. The parameter $\alpha$ and $\beta$ allow
qualitatively for a $p_T$ dependence of the nuclear effect as
observed e.g. in $P+A$ interactions\cite{Antreasyan1,Brenner1}. Now once
more, we come back from the general discussion directly to the
original issue of dealing with and completing the basic approach.
\par
Dividing Eq. (6) for antideuteron production in $AB$ collision by
the square of that for antiproton production in the same collision
one would obtain the expression for $B_2$ with $P_T$ = 2$p_T$, where
$P_T$ denotes the transverse momentum of antideuteron (antiproton).
\section{The Phenomenological Setting for Studying Rapidity-Behaviour of $d(\overline{d})$}
Following Faessler\cite{Faessler1}, Peitzmann\cite{Peitzmann1},
Schmidt and Schukraft\cite{Schmidt1} and finally Thom$\acute{e}$ et
al\cite{Thome1}, we \cite{De2,De5,De3} had formulated in the past a
final working expression for rapidity distributions in proton-proton
collisions at ISR (Intersecting Storage Rings) ranges [$\sqrt{s}$
$\sim$ 20 GeV to 62.4 GeV in center of mass system] of energy-values
by the following three-parameter parametrization, viz,
\begin{equation}
\frac{1}{\sigma}\frac{d\sigma}{dy}=C_1'(1+\exp\frac{y-y_0}{\Delta})^{-1}
\end{equation}
where $C_1'$ is a normalization constant and $y_0$, $\Delta$ are two
parameters. The choice of the above form made by Thom$\acute{e}$ et
al\cite{Thome1} was intended to describe conveniently the central
plateau and the fall-off in the fragmentation region by means of the
parameters $y_0$ and $\Delta$ respectively. Besides, this was based
on the concept of both limiting fragmentation and the Feynman
Scaling hypothesis. For all five energies in $p+p$ collisions the
value of $\Delta$ was obtained to be $\sim$ 0.55 for
pions\cite{De2,De5} and kaons\cite{De3}, $\sim$ 0.35 for
protons/antiprotons\cite{De3}, $\sim$ 0.30 for
deuterons/antideuterons. And these values of $\Delta$ are generally
assumed to remain the same in the ISR ranges of energy. Still, for
very high energies, and for direct fragmentation processes which are
quite feasible in very high energy heavy nucleus-nucleus collisions,
such parameter values do change somewhat prominently, though in most
cases with marginal high energies, we have treated them as nearly
constant.
\par Now, the fits for the rapidity (pseudorapidity)
 spectra for non-pion secondaries produced in the $p+p$ reactions at various energies are phenomenologically
 obtained by De and Bhattacharyya\cite{De3} through the making of
 suitable choices of $C_1'$ and $y_0$. It is observed that for most of the secondaries the values of $y_0$
 do not remain exactly constant and show up some degree of species-dependence .
 However, it gradually increases with energies and the energy-dependence
of $y_0$ is empirically proposed to be expressed by the following
relationship\cite{De2,De5} :

\begin{equation}
y_0=k\ln\sqrt{s_{NN}}+0.8
\end{equation}

\par The nature of energy-dependence of $y_0$ is shown in the adjoining figure
(Fig.1). Admittedly, as k is assumed to vary very slowly with c. m.
energy, the parameter $y_0$ is not exactly linearly correlated to
$\ln \sqrt{s_{NN}}$, especially in the relatively low energy region.
And this is clearly manifested in Fig.1. This variation with energy
in k-values is introduced in order to accommodate and describe the
symmetry in the plots on the rapidity spectra around mid-rapidity.
This is just phenomenologically observed by us, though we cannot
readily provide any physical justification for such perception
and/or observation. And the energy-dependence of $y_0$ is studied
here just for gaining insights in their nature and for purposes of
extrapolation to the various higher energies (in the centre of mass
frame, $\sqrt{s_{NN}}$) for several nucleon-nucleon, nucleon-nucleus
and nucleus-nucleus collisions. The specific energy (in the c.m.
system,
 $\sqrt{s_{NN}}$) for every nucleon-nucleus or nucleus-nucleus collision is first worked out by
 converting the laboratory energy value(s) in the required c.m. frame energy value(s).
 Thereafter the value of $y_0$ to be used for computations of inclusive cross-sections of nucleon-nucleon collisions
 at particular energies of interactions is extracted from Eq. (8) for corresponding obtained energies.
 This procedural step is followed for calculating the rapidity (pseudorapidity)-spectra for not only the pions
 produced in nucleon-nucleus and nucleus-nucleus collisions\cite{De2,De5}. However, for the studies on the rapidity-spectra
 of the non-pion secondaries produced in the same reactions one does always neither have the opportunity to take recourse
 to such a systematic step, nor could they actually resort to this rigorous
 procedure, due to the lack of necessary and systematic data on
 them.

\begin{figure}
\centering
\includegraphics[width=2.5in]{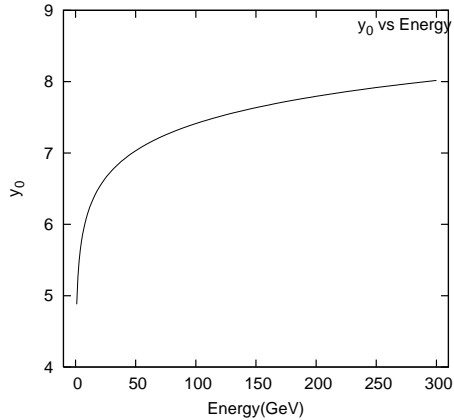}
\caption{Variation of $y_0$ in equation (8) with increasing energy.[Parameter values are shown in Table 1.}
\end{figure}

\par Our next step is to explore the nature of $f(y)$ which is envisaged to be given generally by a polynomial form noted
below :
\begin{equation}
f(y) = \alpha' + \beta' y + \gamma' y^2,
\end{equation}
where $\alpha'$, $\beta'$ and $\gamma'$ are the coefficients to be
chosen separately for each AB collisions (and also for AA collisions
when the projectile and the target are same). Besides, some other
points are to be made here. The suggested choice of form in
expression (9) is not altogether fortuitous. In fact, we got the
clue from one of the previous work by one of the authors
(SB)\cite{Bhattacharyya1} here pertaining to the studies on the
behavior of the EMC effect related to the lepto-nuclear collisions.
In the recent past Hwa et al.\cite{Hwa1} also made use of this sort
of relation in a somewhat different context. Now let us revert to
our original discussion and to the final working formula for
$\frac{dN}{dy}$ in various A+B (or A+A) collisions given by the
following relation :
\begin{equation}
\frac{dN}{dy}|_{AB \rightarrow QX} = C_2'(AB)^{\alpha' + \beta' y + \gamma' y^2}\frac{dN}{dy}|_{PP \rightarrow QX}
 = C_3'(AB)^{\beta' y + \gamma' y^2}(1 + \exp \frac{y-y_0}{\Delta})^{-1},
\end{equation}
where $C_2'$ is the normalization constant and
$C_3'$=$C_2'(AB)^{\alpha'}$ is another constant as $\alpha'$ is also
a constant for a specific collision at a specific energy.
\section{Results}
Most of the results obtained on the basis of the models applied here
are presented diagrammatically. The theoretical results on inclusive
cross-sections for production of deuteron, antideuteron, proton and
antiproton production in $Pb+Pb$ collision at 158A GeV are displayed
in Fig.2 and Fig.3. The solid curves in Fig.2 and Fig.3 are the
presentations of the remarkably good production of the inclusive
cross-section and the solid lines are the depictions of the results
obtained on the basis of expression (6). The necessary parameter
values related to these four figures are given in Table 2 to Table
5. The theoretical plots are made against the data-sets on
normalized inclusive cross-sections versus $p_T$ for production of
deuteron, antideuteron, proton and antiproton. The DBP-based
calculations show somewhat fair agreement with measured data in all
these cases. These plots are being presented here in order to make a
few points as particular observations : (i) The ratio-values
obtained theoretically have a substantial support from the data on
the basic observables. (ii) The agreement between model-based
calculations and data is unlikely to be fortuitous, as it is found
in such widely varied particle-species of secondaries like the
deuteron, antideuteron, proton and antiproton. (iii) The testing is
done here for production of $d$ and $\overline{d}$ in collisions
with the simplest of projectile and target ($p+p$) and also in
collisions involving the heaviest nuclei like lead-lead. The other
rapidity dependence behaviour studies for $d$ and $\overline{d}$
production in $Au+Au$ and $Si+Pb$ at different energies was made by
De and Bhattacharyya\cite{De4}. This exposes modestly the wide range
of applicability of the studied approach.

\par
The four important ratio behaviours are shown in Fig.4 and Fig.5.
Fig.4 shows the ratio between $\overline{d}/d$ and $\overline{p}/p$
and Fig.5 shows the ratio between $\overline{p}/\overline{d}$ and
$p/d$. The coalescence parameters for $Pb+Pb$ collisions are plotted
in Fig.6 and in Fig.7 respectively.
\par
Some comments on the rapidity-spectra are now in order. Here we draw
the rapidity-density of deuteron and antideuteron for symmetric
$Pb+Pb$ collisions at 158A GeV which have been appropriately labeled
at the top right corner of Fig.8. Though the figure represents the
case for production of deuteron and antideuteron, we fail to
understand physically why the data depict exactly opposite nature of
$\frac{dN}{dy}$ dependence of $d$ and $\overline{d}$. Besides, the
solid curves in all cases-almost without any exception-demonstrate
our GCM-based results. Secondly, the data on rapidity-spectra for
some high-energy collisions are, at times, available for both
positive and negative y-values. This would give rise to a problem in
our method for studying the asymmetric collisions wherein the
colliding nuclei are of non-identical nature. This is because, in
our expression (10) the coefficient $\beta'$ multiplies a term which
is proportional to y and so is not symmetric under
y$\rightarrow$(-y). In order to overcome this difficulty we would
introduce here $\beta'$=0 for all the graphical plots. And for
symmetric central collisions this is not an unphysical proposition
or assumption. Of course, for mid-central collisions (as the case
here is), we are afraid, some divergence with data-behaviour might
arise. These plots are represented by Fig.8 for deuteron and
antideuteron in $Pb+Pb$ collisions. The parameter values in this
particular case are presented in Table 6. The diagrams shown in
Fig.9 represent the model-based results on $\overline{d}/d$. These
plots are drawn on the basis of the figures shown in Fig.8 with the
fit-parameters given in Table 6. The data-trends have been captured
by our plot in a modestly right way, though it is at variance with
what had been shown to be the nature of the data by the dotted
straight line shown in Fig.5 by NA49 Collaboration\cite{Anticic1}.

\section{Concluding Remarks}
The approach adopted here, though new, is just purely
phenomenological. Still, the method reproduces the data-trends on
the $p_T$ and $y$ -spectra on $d$ and $\overline{d}$ production in
the studied collision with a fair degree of success. Some predictive
plot(s) in our previous work might appear to be at variance with the
plots presented here. But it is to be borne in mind that the
previous measurements were not centrality-based studies for which
such differences between the behaviour of the past and present,
data-sets are quite possible and cannot be treated as such as an
`anomaly'. However, as data on helium-antihelium and
triton-antitriton are still too sparse and they suffer from large
error-bars, we have chosen not to focus on them in the present
context. Our model-based values (Fig. 7) of the coalescence factor
(parameters) for $d$ and $\overline{d}$ are somewhat less than the
values depicted by pure coalescence model. The obvious implification
of this is two-fold : Either the models that have been made use of
here are not perfectly alright or the standard coalescence picture
here has got to be modified. Still, unless the highly reliable data
on deuteron, antideuteron, triton-antitriton are available, we
cannot make any definitive conclusion(s) on the validity (or the
lack of it), in so far as the involved physical ideas are concerned.
\par
However, on the whole, it is quite striking to note that
rapidity-density yields for $d$ and $\overline{d}$ are qualitatively
and quantitatively in fair agreement with the data-trends measured
by NA49 collaboration. And these data-behaviours are obtained just
by the chosen pattern of expressions and not by any other
mathematical trick. Besides, we obtain the ratio of rapidity-density
of $d$ and $\overline{d}$ for $Pb+Pb$ collisions at 158A ($\sqrt{s}$
$\approx$ 17.3) GeV to be $\sim$ 3.2 $\times$10$^{-3}$ and this
ratio is to be checked by the future experiments. Furthermore, the
$p_T$-dependence nature, $y$-dependence characteristics of
$\overline{d}/d$ ratio behaviours demonstrated by our model are, by
look, almost similar. These have also to be scrutinized from the
future studies.
\par
In recent times, preliminary data on cosmic antideuteron
flux\cite{Ibarra1,Vittino1} have just arrived. Further data on
cosmic $d$ and $\overline{d}$ from various experiments (The p-GAPS
experiment)\cite{Mognet1,Fuke1} are expected to be available by
2017/2018. When reliable high-statistics cosmic ray data would be at
hand, we would try to deal with them at appropriate time.
\par
Despite the few successes indicated above, the main deficiencies of
the present approach cannot and should not be overlooked. In our
approach, there are some free parameters which must be physically
identified. In otherwords, the parameters $\alpha$, $\beta$,
$\alpha'$, $\beta'$ etc will have to be interpreted in terms of the
obsevables of physics of collisions, like impact parameter,
centrality of the collisions, number of
participant-nucleons/constituents and the c. m. energy of the
colliding systems etc. In order to do so what we need is more
high-statistics reliable data on the measured observables.
\newpage
\begin{center}
\par{\textbf{Acknowledgements}}
\end{center}
\par
The authors are grateful to the two learned referees for encouraging
remarks, helpful comments and some valuable suggestions.

\newpage
{\singlespacing{
\begin{table}
\begin{center}
\begin{small}
\caption{Variation of $y_0$ with Energy.}
\begin{tabular}{|c|c|c|c|}\hline
 $Energy (A GeV)$& $\sqrt{s_{NN}}(GeV)$ & $Constant (k)$ & $y_0$ \\
 \hline
 $20$& $6.3$ & $ 2.76$ & $5.894$ \\
 \hline
 $30$& $7.6$ & $2.54$ & $6.006$\\
 \hline
 $40$& $8.7$ & $2.40$ & $6.085$\\
 \hline
 $80$& $12.3$ & $2.16$ & $6.276$\\
 \hline
 $158$& $17.3$ & $1.97$ & $6.463$\\
 \hline
\end{tabular}
\end{small}
\end{center}
\end{table}

\begin{table}
\begin{center}
\begin{small}
\caption{Different parameter values for deuteron production in
$Pb+Pb$ collisions at 158A GeV}
\begin{tabular}{|c|c|c|c|c|c|c|}\hline
 $Centrality$& $C_3$ & $\alpha$ & $\beta$ & $p_0$ & $n$ & $\frac{\chi^2}{ndf}$\\
 \hline
 $0-12.5\%$& $0.363\pm0.003$ & $0.563\pm0.001$ & $0.099\pm0.001$ & $4.366\pm0.008$ & $26.664\pm0.043$ & $1.675/10$ \\
 \hline
 $12.5\%-23.5\%$& $0.272\pm0.003$ & $0.500\pm0.001$ & $0.097\pm0.001$ & $4.996\pm0.016$ & $26.236\pm0.075$ & $2.877/10$ \\
 \hline
 $0-23.5\%$& $0.319\pm0.002$ & $0.563\pm0.001$ & $0.102\pm0.001$ & $4.366\pm0.009$ & $26.331\pm0.049$ & $2.032/10$ \\
 \hline
\end{tabular}
\end{small}
\end{center}
\end{table}

\begin{table}
\begin{center}
\begin{small}
\caption{Different parameter values for antideuteron production in
$Pb+Pb$ collisions at 158A GeV}
\begin{tabular}{|c|c|c|c|c|c|c|}\hline
 $Centrality$& $C_3$ & $\alpha$ & $\beta$ & $p_0$ & $n$ & $\frac{\chi^2}{ndf}$\\
 \hline
 $0-12.5\%$& $0.0008\pm6.098\times10^{-6}$ & $0.215\pm0.001$ & $0.136\pm0.001$ & $13.300\pm0.104$ & $21.307\pm0.162$ & $1.824/03$ \\
 \hline
 $12.5\%-23.5\%$& $0.0009\pm1.716\times10^{-5}$ & $0.113\pm0.002$ & $0.050\pm0.002$ & $11.999\pm0.151$ & $22.942\pm0.278$ & $1.068/03$ \\
 \hline
 $0-23.5\%$& $0.0009\pm4.613\times10^{-6}$ & $0.157\pm0.001$ & $0.055\pm0.001$ & $11.415\pm0.043$ & $24.359\pm0.089$ & $0.525/03$ \\
 \hline
\end{tabular}
\end{small}
\end{center}
\end{table}

\begin{table}
\begin{center}
\begin{small}
\caption{Different parameter values for proton production in $Pb+Pb$
collisions at 158A GeV}
\begin{tabular}{|c|c|c|c|c|c|c|}\hline
 $Centrality$& $C_3$ & $\alpha$ & $\beta$ & $p_0$ & $n$ & $\frac{\chi^2}{ndf}$\\
 \hline
 $0-12.5\%$& $51.669\pm0.233$ & $0.338\pm0.001$ & $0.139\pm0.001$ & $4.985\pm0.010$ & $17.932\pm0.031$ & $12.035/11$ \\
 \hline
 $12.5\%-23.5\%$& $42.023\pm0.201$ & $0.321\pm0.001$ & $0.134\pm0.001$ & $4.000\pm0.007$ & $15.017\pm0.024$ & $13.810/10$ \\
 \hline
 $0-23.5\%$& $51.937\pm0.260$ & $0.309\pm0.001$ & $0.127\pm0.001$ & $5.000\pm0.009$ & $18.001\pm0.028$ & $11.064/10$ \\
 \hline
\end{tabular}
\end{small}
\end{center}

\begin{center}
\begin{small}
\caption{Different parameter values for antiproton production in
$Pb+Pb$ collisions at 158A GeV}
\begin{tabular}{|c|c|c|c|c|c|c|}\hline
 $Centrality$& $C_3$ & $\alpha$ & $\beta$ & $p_0$ & $n$ & $\frac{\chi^2}{ndf}$\\
 \hline
 $0-12.5\%$& $5.092\pm0.031$ & $0.400\pm0.001$ & $0.123\pm0.001$ & $5.000\pm0.012$ & $24.750\pm0.054$ & $3.869/14$ \\
 \hline
 $12.5\%-23.5\%$& $3.733\pm0.028$ & $0.320\pm0.001$ & $0.120\pm0.001$ & $4.923\pm0.017$ & $20.004\pm0.066$ & $8.483/14$ \\
 \hline
 $0-23.5\%$& $4.460\pm0.024$ & $0.350\pm0.001$ & $0.122\pm0.001$ & $4.785\pm0.011$ & $21.000\pm0.046$ & $3.495/14$ \\
 \hline
\end{tabular}
\end{small}
\end{center}

\begin{center}
\begin{small}
\caption{Values of different parameters for production of deuteron
and antideuteron in the 23.5\% most central $Pb+Pb$ collisions at
158A GeV (for $\beta'$=0) for both +ve and -ve rapidities.}
\begin{tabular}{|c|c|c|c|}\hline
  $Production$ & $C_3'$ & $\gamma'$ & $\frac{\chi^2}{ndf}$\\
 \hline
 $d$ & $0.242\pm2.70\times10^{-4} $  & $0.034 \pm0.0001$ & $0.160/03 $ \\
 \hline
  $\overline{d}$ & $0.001\pm1.00\times10^{-6} $ & $-0.050 \pm0.0001 $& $0.016/03 $ \\
 \hline
 \end{tabular}
\end{small}
\end{center}
\end{table}

\newpage

\begin{figure}
\subfigure[]{
\begin{minipage}{.5\textwidth}
\centering
\includegraphics[width=2.5in]{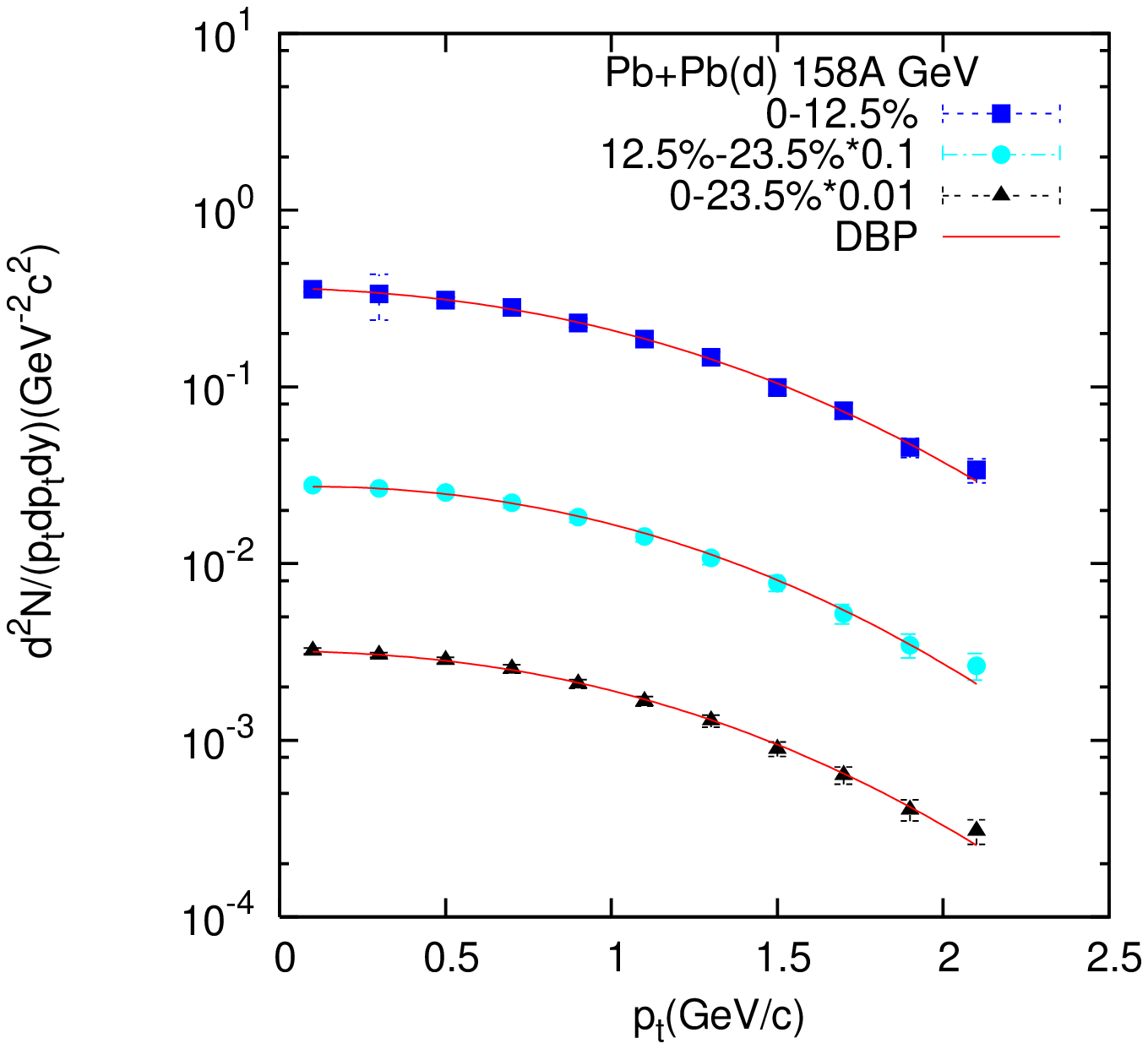}
\end{minipage}}%
\subfigure[]{
\begin{minipage}{.5\textwidth}
\centering
 \includegraphics[width=2.5in]{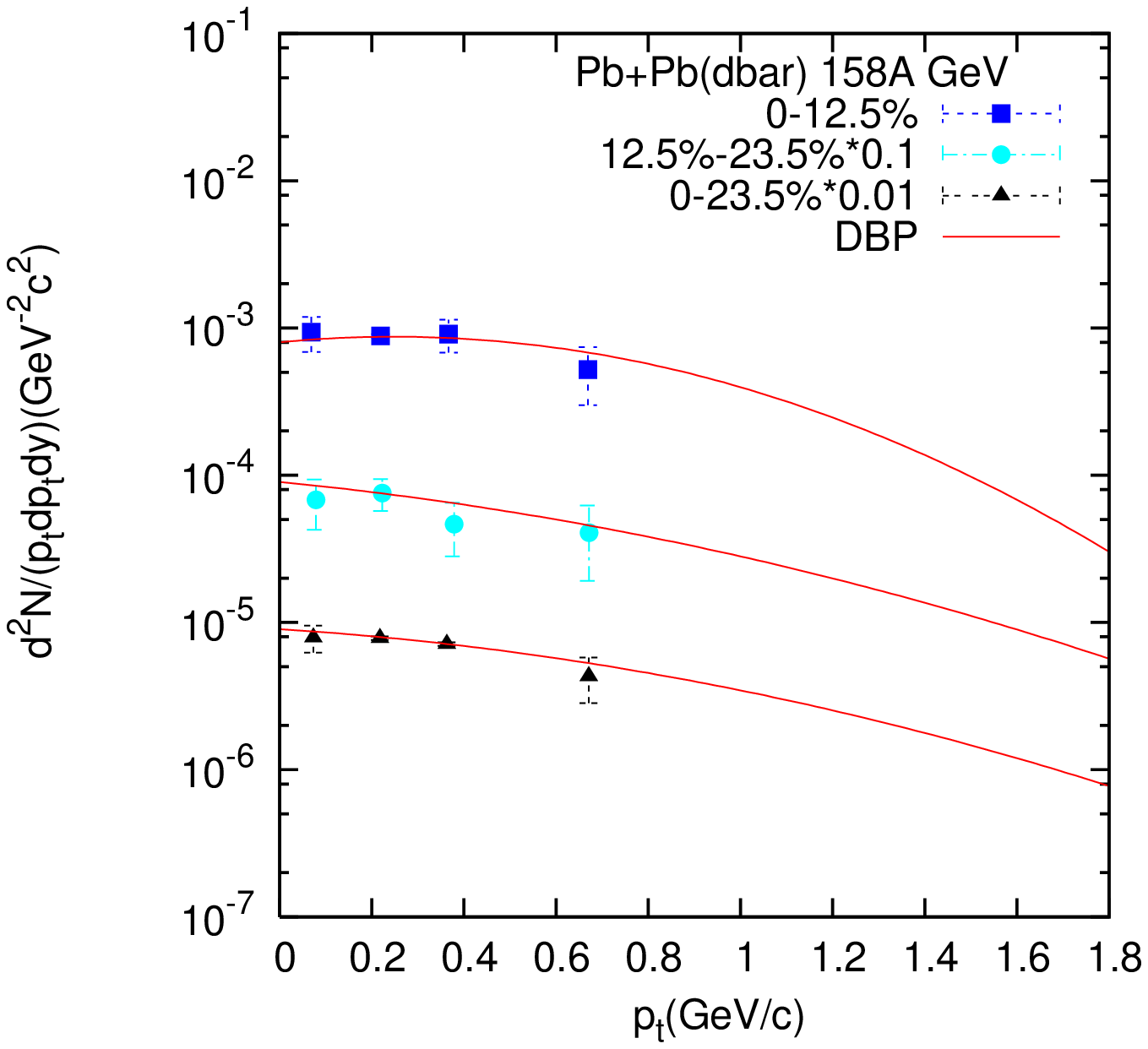}
 \end{minipage}}%
\caption{The transverse momentum spectra of deuterons (a) antideuteron (b) in centrality selected $Pb+Pb$
 collisions at 158A GeV. Only statistical errors are shown. The experimental data are taken from {\cite{Anticic1}} and the parameter
values are taken from Table 2 \& Table 3. The model-based results of the present work are shown
by the solid curves.}
\end{figure}

\begin{figure}
\subfigure[]{
\begin{minipage}{.5\textwidth}
\centering
\includegraphics[width=2.5in]{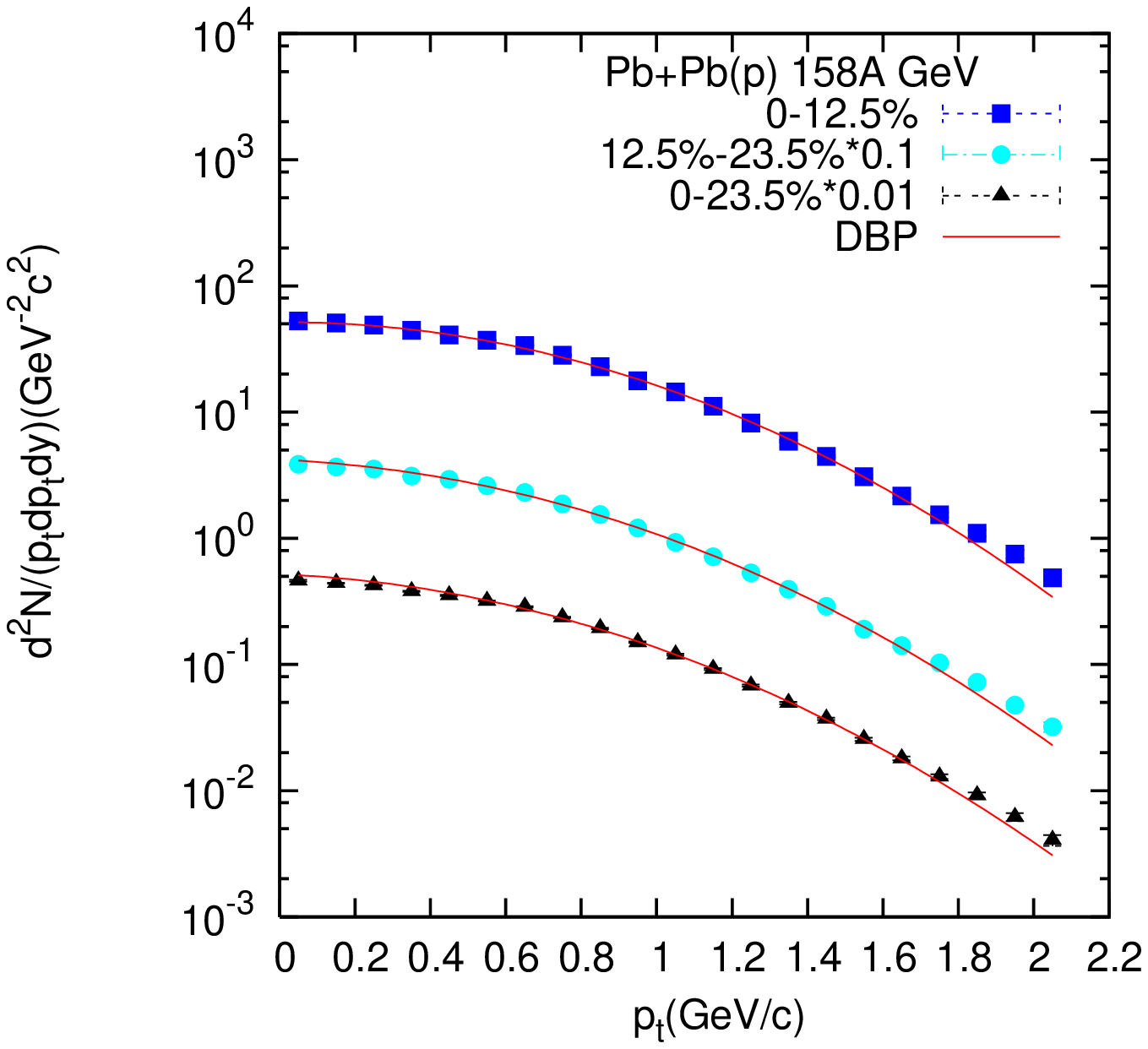}
\end{minipage}}%
\subfigure[]{
\begin{minipage}{.5\textwidth}
\centering
 \includegraphics[width=2.5in]{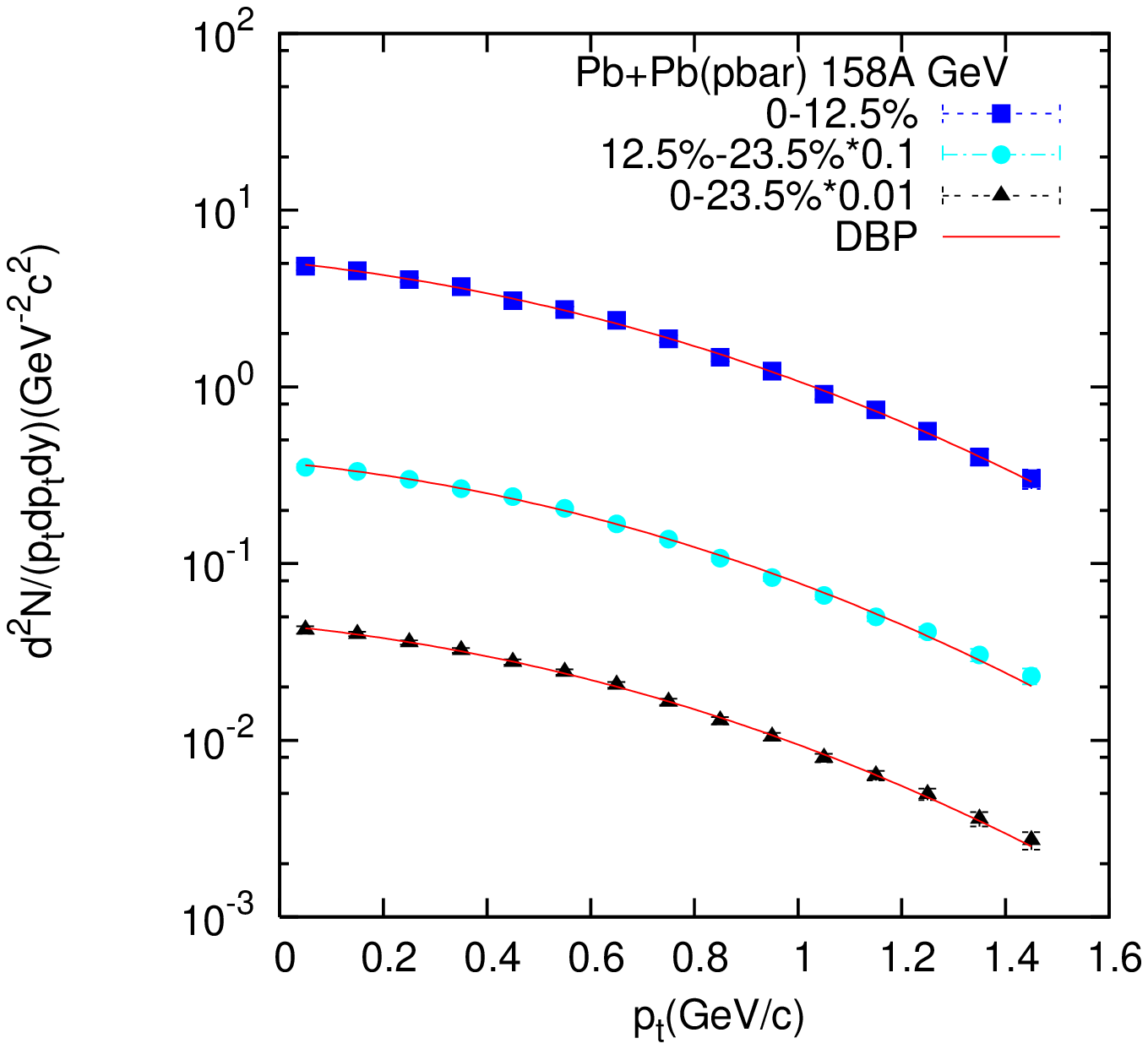}
 \end{minipage}}%
\caption{The transverse momentum spectra of protons (a) antiprotons (b) in centrality selected $Pb+Pb$
 collisions at 158A GeV. Only statistical errors are shown. The experimental data are taken from {\cite{Anticic1}} and the parameter
values are taken from Table 4 \& Table 5. The solid curves provide the present model-based results.}
\end{figure}

\begin{figure}
\subfigure[]{
\begin{minipage}{.5\textwidth}
\centering
\includegraphics[width=2.5in]{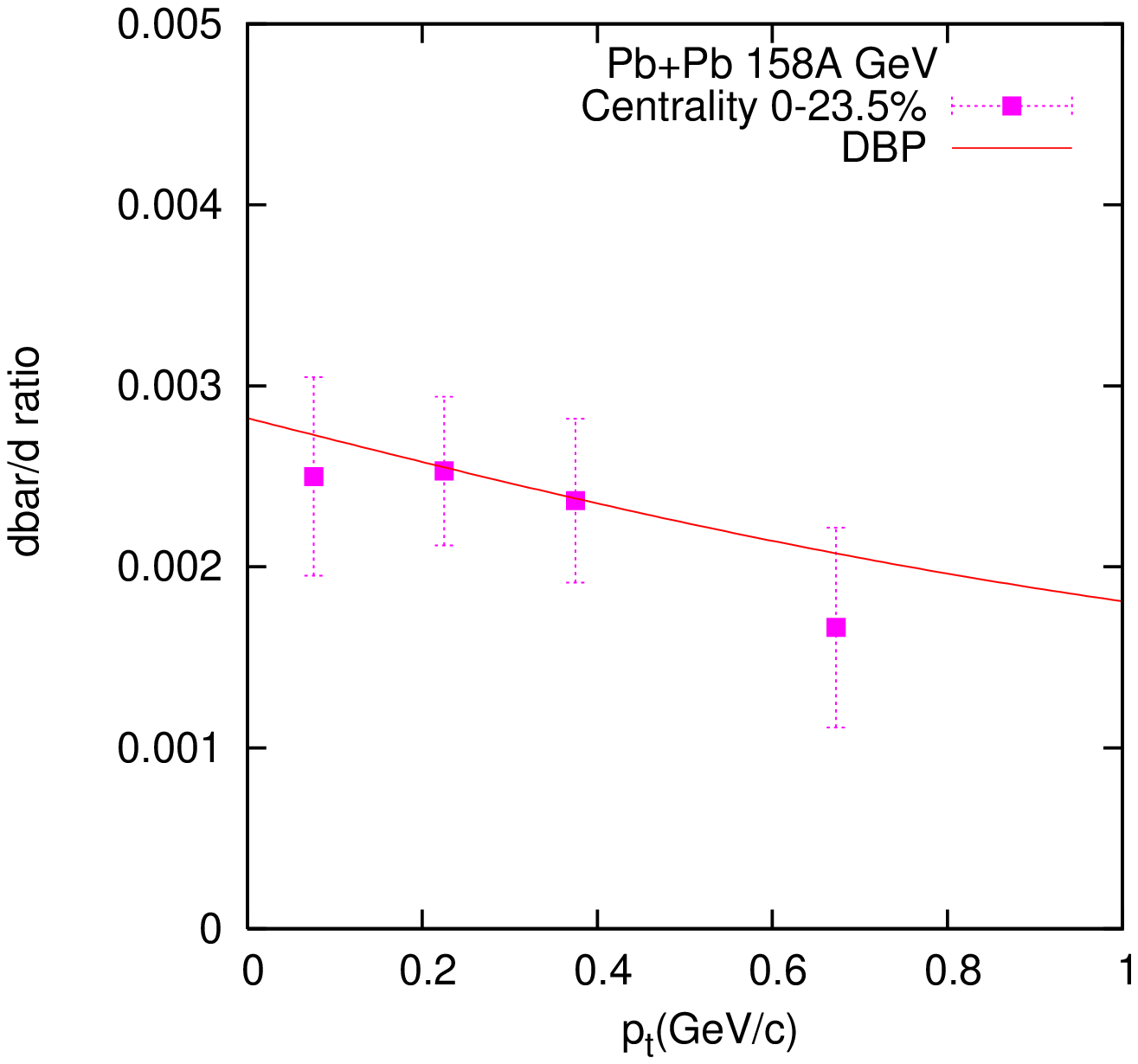}
\end{minipage}}%
\subfigure[]{
\begin{minipage}{.5\textwidth}
\centering
 \includegraphics[width=2.5in]{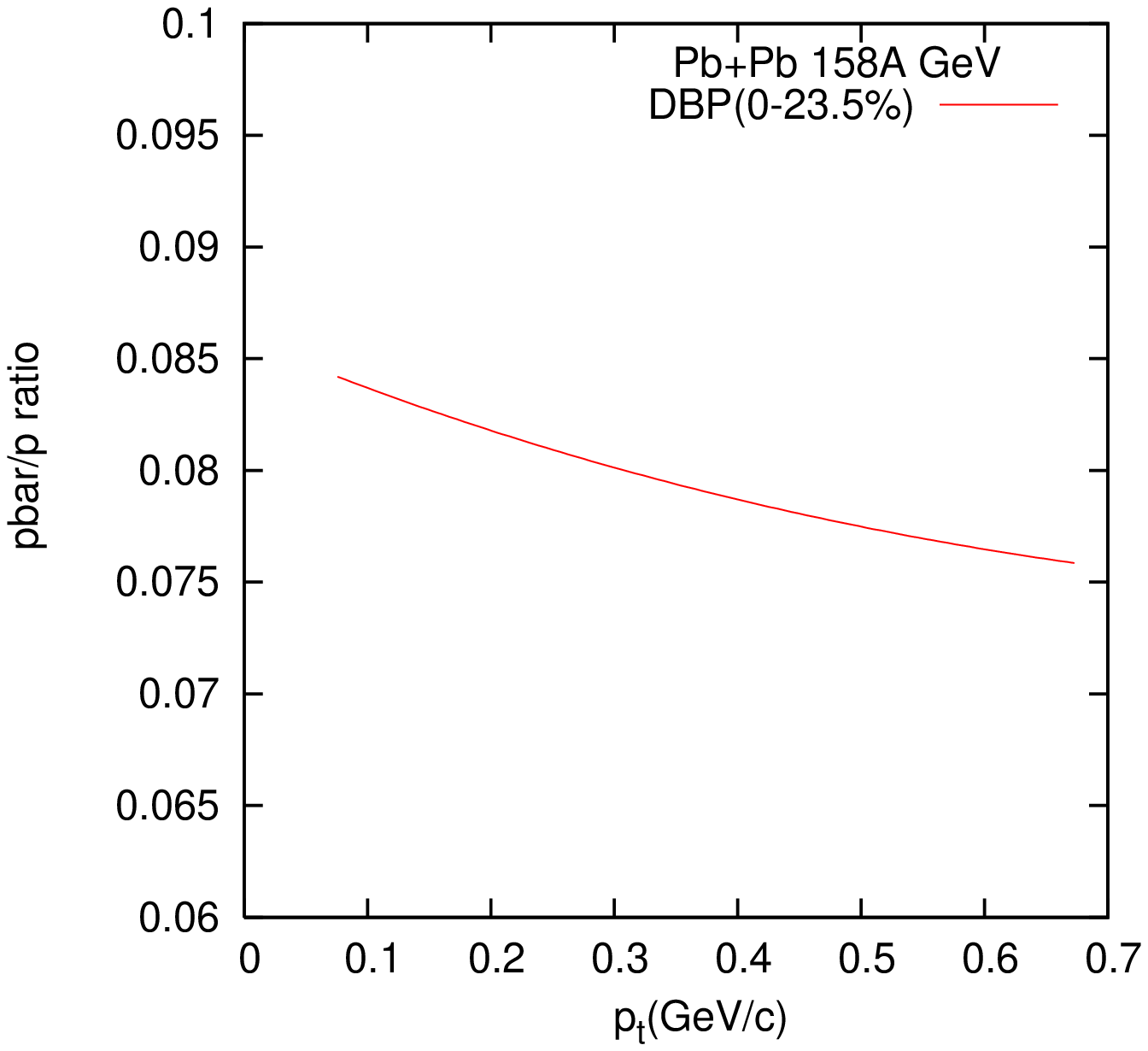}
 \end{minipage}}%
\caption{(a)$\overline{d}/d$-ratio  as a function of $p_T$ in 0-23.5\% central Pb+Pb collisions at 158A GeV and
 (b)Predictive nature of $\overline{p}/p$-ratio with respect to the transverse
momentum ($p_T$) in 0-23.5\% central $Pb+Pb$ Collisions at 158A GeV on the basis of present model.
The experimental data are taken from {\cite{Anticic1}}.}
\end{figure}

\begin{figure}
\subfigure[]{
\begin{minipage}{.5\textwidth}
\centering
\includegraphics[width=2.5in]{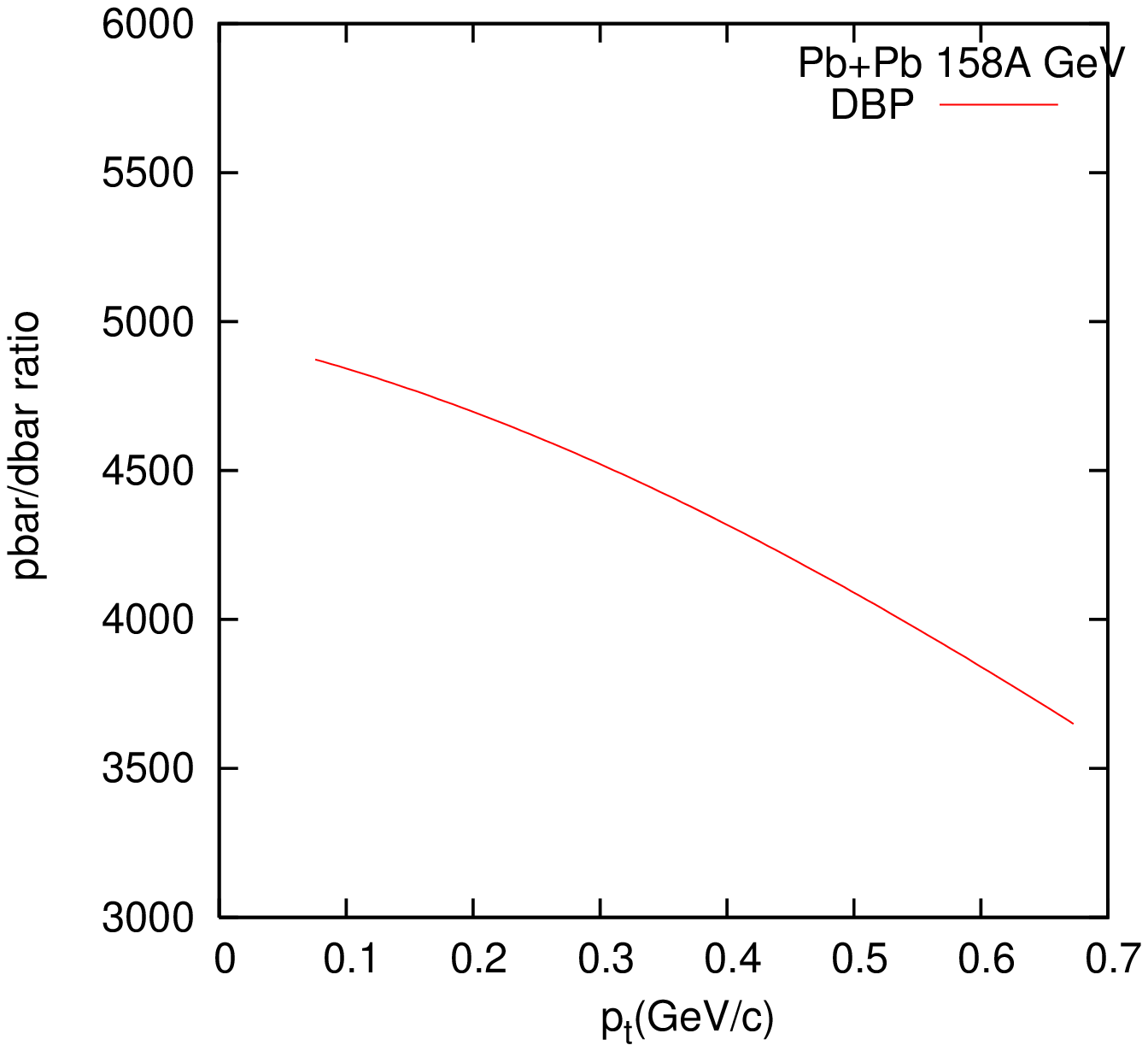}
\end{minipage}}%
\subfigure[]{
\begin{minipage}{.5\textwidth}
\centering
 \includegraphics[width=2.5in]{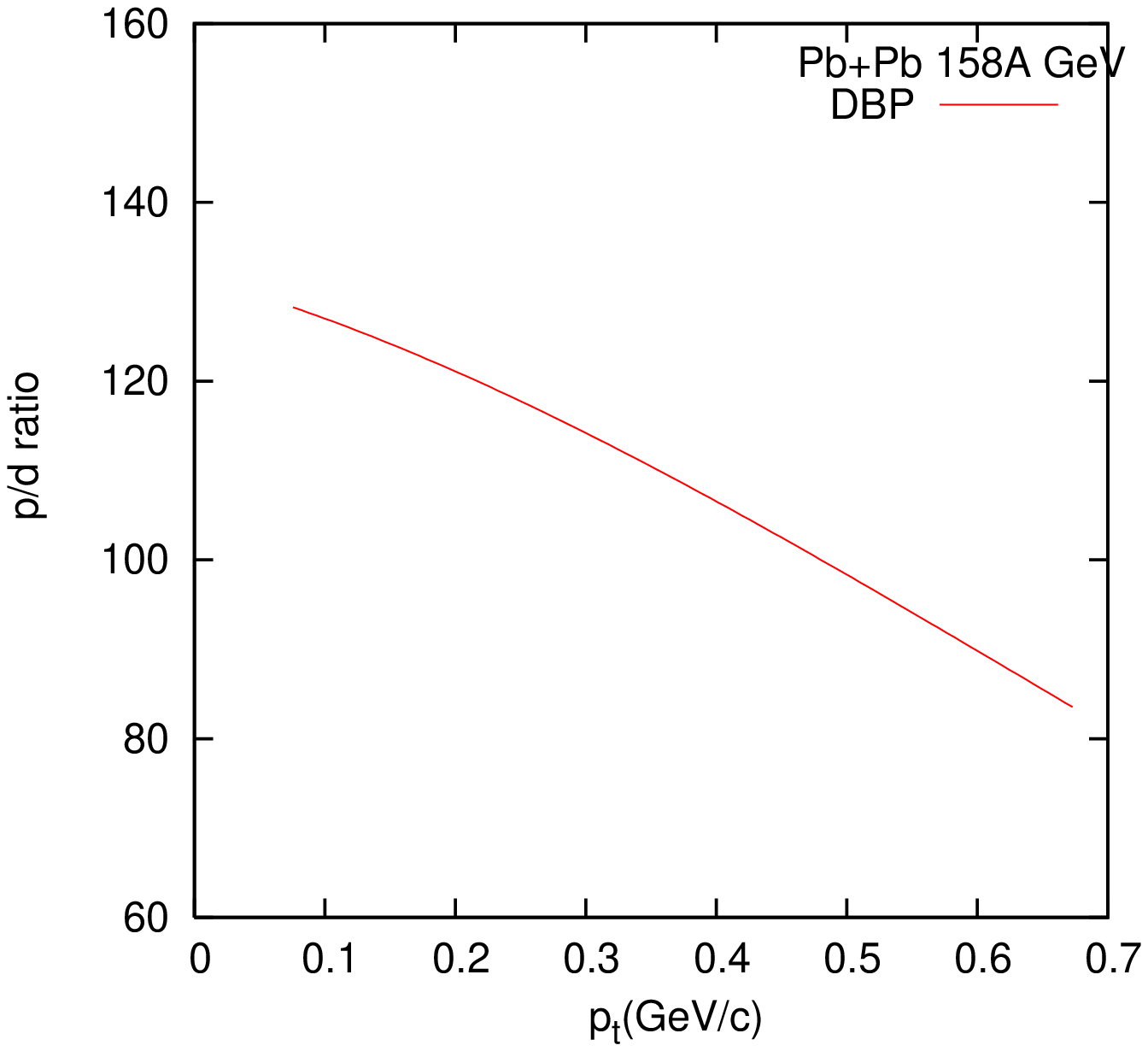}
 \end{minipage}}%
\caption{Predictive nature of $\overline{p}/\overline{d}$-ratio and $p/d$-ratio with respect to the transverse
momentum ($p_T$) in 0-23.5\% central $Pb+Pb$ Collisions at 158A GeV on the basis of present model.}
\end{figure}

\begin{figure}
\centering
\includegraphics[width=2.5in]{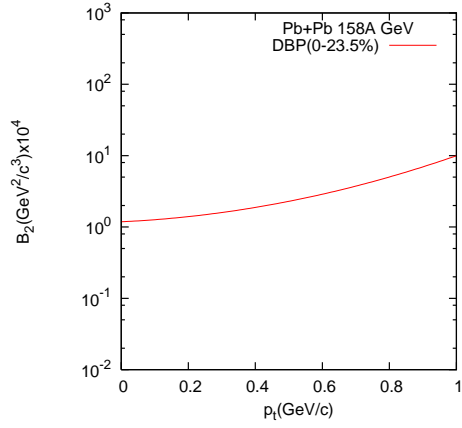}
\caption{Plot of Coalescence parameter $B_2$ as a function of antideuteron transverse momentum
$p_T$ in the case of $Pb+Pb$ collision. The solid curve depicts the DBP-based results.}
\end{figure}

\begin{figure}
\centering
\includegraphics[width=2.5in]{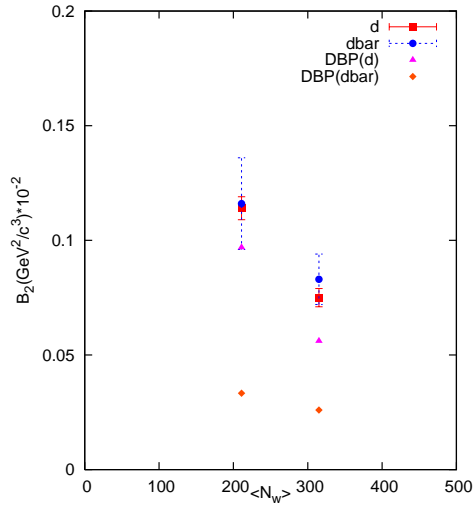}
\caption{Plot of Coalescence parameter $B_2$ for $d$ and $\overline{d}$ calculated from
DBP-model in centrality selected $Pb+Pb$ collision at 158A GeV.}
\end{figure}

\begin{figure}
\centering
\includegraphics[width=2.5in]{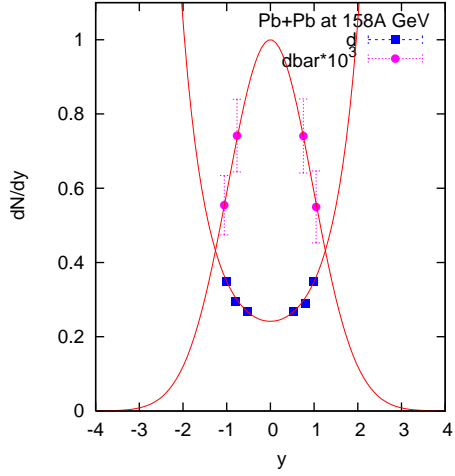}
\caption{Rapidity-density distributions for $\overline{d}$ (circles) and $d$ (squares) produced in the 23.5 \%
most central $Pb+Pb$ collisions at 158A GeV. The symbols are the experimental data and the data
points are taken from {\cite{Anticic1}} and the parameter
values are taken from Table 6. The solid curve provide the GCM-based results.}
\end{figure}

\begin{figure}
\centering
\includegraphics[width=2.5in]{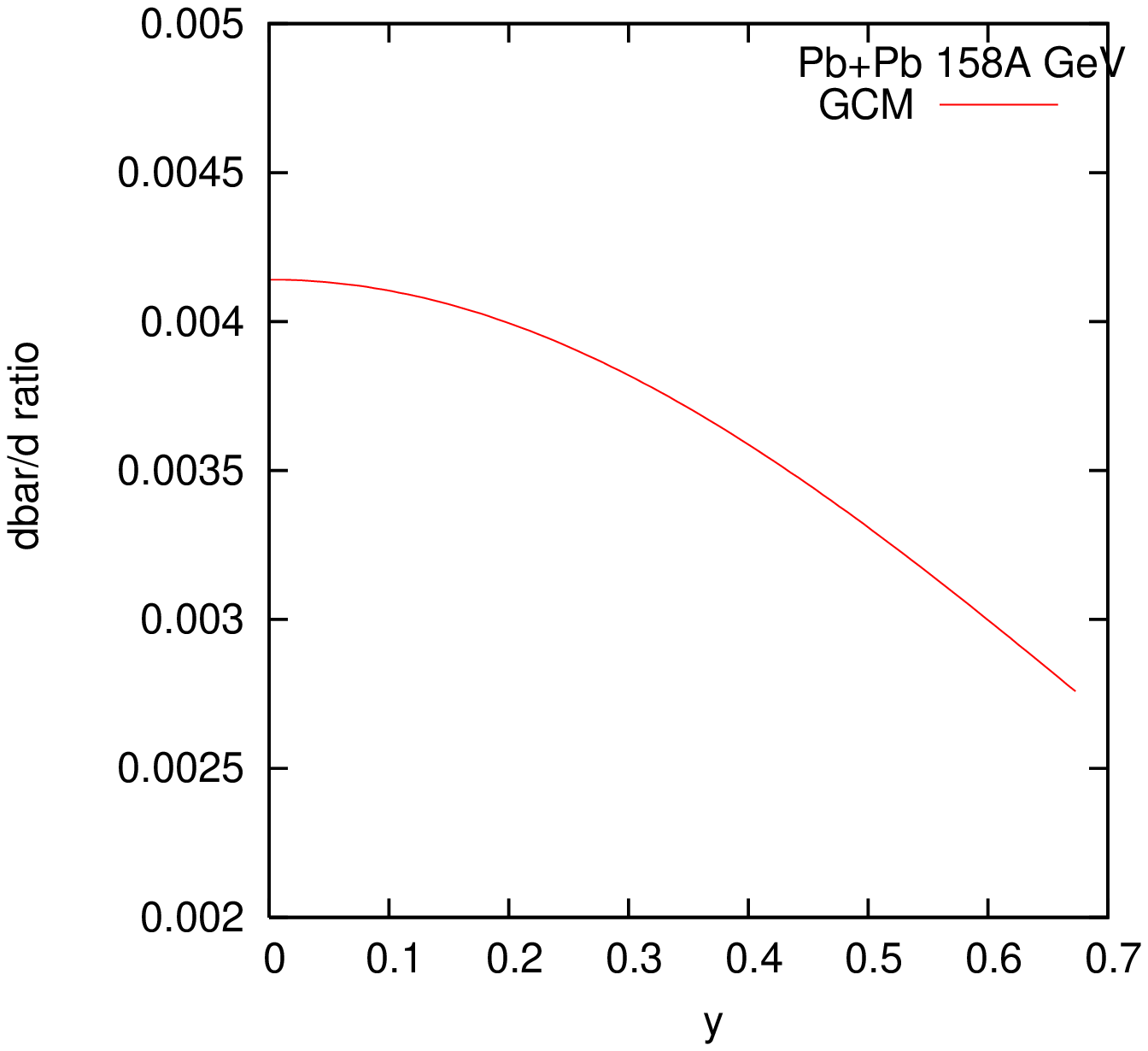}
\caption{Predictive nature of $\overline{d}/d$-ratio with respect to the rapidity (y) in 23.5\% central $Pb+Pb$ Collisions at 158A GeV on the basis of
GCM-based results.}
\end{figure}

\begin{thebibliography}{*}
\bibitem{Anticic1} T. Anticic et al. (NA49 Collaboration), Phys.
Rev. \textbf{C 85}, 044913 (2012) [nucl-ex/1111.2588 v1 10 November
2011] and the references therein.
\bibitem{Kolesnikov1} V. I. Kolesnikov et al. (NA49 Collaboration), J. Phys.
Conf. Ser. \textbf{110}, 032010 (2008) and the references therein.
\bibitem{Gang1}C. Gang, C. Huan, W. Jiang-Ling and C. Zheng-Yu : hep-ph/1401.6872 v1 27 Jan. 2014
and the references therein.
 \bibitem{Ma1}Y. G. Ma, J. H. Chen and L. Xue, Frontiers of Physics
 (Springer Publication) \textbf{7(6)}, 637 (2012).
\bibitem{Shuryak1}E. Shuryak, Nuclei and Atoms \textbf{23}, 501 (2010).
\bibitem{Shuryak2}E. Shuryak, Prog. Part. Nucl. Phys. \textbf{62}, 48 (2009).
  \bibitem{Blaizot1}J. P. Blaizot, "INT $@$ 20 : The Future of Nuclear Physics and its
  Intersection", INT, Seatle (Washington), July 2010.
 \bibitem{Satz1}H. Satz, Nucl. Phys. \textbf{A 862-863}, 4 (2011).
\bibitem{Tannenbaum1}M. J. Tannenbaum, QNP 2012, Quark and Nuclear Physics,
April 16-20, 2012, Ecole Polytechnique.

 \bibitem{Ellis1}J. Ellis et al., Phys. Lett. \textbf{B 233}, 223 (1989).
 \bibitem{Heinz1}U. Heinz et al., J. Phys. \textbf{G 12}, 1237
 (1989).
 \bibitem{Canetti1}L. Canetti, M. Drewes and M. Shaposhnikov, New. Jour.
 Phys. \textbf{14}, 095012 (2012) and the references therein.
 \bibitem{Gonzales1}R. Gonzales Felife, Int. Jour. Mod. Phys. \textbf{E 20}, supp 01,
 56 (2011).
 \bibitem{Wikzek1}F. Wikzek, Sc. America \textbf{243}, 82 (1980).
 \bibitem{Basini1}G. Basini, A. Morselli and M. Ricci, La Riv Del Nuovo
 Cimento 12(4), 1 (1989).
\bibitem{Hagedorn1} R. Hagedorn, Riv. Nuovo. Cimento \textbf{6}, 10 (1983).
\bibitem{De1} B. De and S. Bhattacharyya, Eur. Phys. Jour. \textbf{A 10}, 387
(2001).
\bibitem{Bhattacharyya2}S. Bhattacharyya and B. De, Mod. Phys. Lett.
\textbf{A 16}, 1395 (2001).
\bibitem{Dasgupta1}S. Dasgupta ana A. Z. Mekjian, Phys. Rep. \textbf{72}, 131 (1981).
\bibitem{Mekjian1}A. Z. Mekjian, Phys. Rev. \textbf{C 17}, 1051 (1978).
\bibitem{Peitzmann1}T. Peitzmann, Phys. Lett. \textbf{B 450}, 7 (1999).
\bibitem{De2}B. De, S. Bhattacharyya and P. Guptaroy, Int. J. Mod. Phys. \textbf{A 17}, 4615
(2002).
\bibitem{De5}B. De, S. Bhattacharyya and P. Guptaroy, Jour.
Phys. \textbf{G 28}, 2963 (2002).
\bibitem{Bhattacharyya1}S. Bhattacharyya, Lett. Nuvo. Cim. \textbf{44(2)}, 119 (1985).
\bibitem{Aubert1}J. Aubert et al. (The EMC collaboration), Phys. Letts. \textbf{B 123}, 275 (1983).
\bibitem{Bodek1}A. Bodek et al., Phys. Rev. Letts. \textbf{50}, 1431 (1983).
\bibitem{Hwa1}R. C. Hwa et al., Phys. Rev. \textbf{C 64}, 054611 (2001).
\bibitem{VanHove1}L. Van-Hove, Z. Phys. \textbf{C 21}, 93 (1983).
\bibitem{Antreasyan1}D. Antreasyan et al., Phys. Rev. \textbf{D 19}, 764 (1979).
\bibitem{Brenner1}A. E. Brenner et al., Preprint, FERMILAB-Conf-80/47-EXP(June 1980).
\bibitem{Faessler1} M. A. Faessler, Phys. Rep. \textbf{115}, 1 (1984).
\bibitem{Schmidt1}H. R. Schmidt and J. Schukraft, J. Phys. \textbf{G 19}, 1705
(1993).
\bibitem{Thome1}W.Thom$\acute{e}$ et al., Nucl. Phys. \textbf{B 129}, 365 (1977).
\bibitem{De3}B. De and S. Bhattacharyya, Int. J. Mod.Phys. \textbf{A 19},
2313 (2004).
\bibitem{De4}B. De and S. Bhattacharyya, Fizika \textbf{B 13},
665 (2004).
\bibitem{Ibarra1}I. Ibarra and S. Wild, Phys.
Rev. \textbf{D 88}, 023014 (2013) [astro-ph.HE/1301.3820 v1 16
January 2013] and the references therein.
\bibitem{Vittino1}A. Vittino, N. Fornengo and  L. Maccione, Proceedings of the RICAP 2013
Conference, Roma, Italy, May 22-24, 2013 [hep-ph/1308.4848 v1 22
August 2013].
\bibitem{Mognet1}S. A. I. Mognet et al. (The Prototype GAPS Experiment) :
asro-ph.IM/1303.1615 v1 7 March 2013.
\bibitem{Fuke1}H. Fuke et al. (The Prototype GAPS Experiment) : astro-ph.IM/1303.0380 v3 24 June 2013
[Accepted for publication in  Advances in Space Research]


\end{thebibliography}
\end{document}